\documentstyle[prl,aps,multicol,epsf]{revtex}
\renewcommand{\narrowtext}{\begin{multicols}{2}
\global\columnwidth20.5pc} 
\renewcommand{\widetext}{\end{multicols}
\global\columnwidth42.5pc} \multicolsep = 8pt plus 4pt minus 3pt

\newcommand{\be}{\begin{equation}}
\newcommand{\ee}{\end{equation}}

\begin{document}
\draft
\title{Single Electron Transport in Ropes of Carbon Nanotubes}

\author{Marc Bockrath, David H. Cobden, Paul L. McEuen, Nasreen G.
Chopra, and A. Zettl}
\address{Molecular Design Institute, Lawrence Berkeley National
Laboratory and\\ Department of Physics, University of California at
Berkeley}

\author{Andreas Thess, and R. E. Smalley}
\address{ Molecular Design Institute, Lawrence Berkeley National
Laboratory and \\Department of Chemistry, Rice University}

\date{\today} 
\maketitle

\begin{abstract}
We have measured the electrical properties of individual bundles, or
"ropes" of single-walled carbon nanotubes. Below ~10 K, the low bias
conductance is suppressed for voltages below a few millivolts.  In
addition, dramatic peaks are observed in the conductance as a function of
a gate voltage that modulates the number of electrons in the rope.  We
interpret these results in terms of single electron charging and resonant
tunneling through the quantized energy levels of the nanotubes comprising
the rope. 
\end{abstract}

\narrowtext

In the last decade, transport measurements have emerged as a primary tool
for exploring the properties of nanometer-scale structures.  For example,
studies of quantum dots have illustrated that single electron charging and
resonant tunneling through quantized energy levels regulate transport
through small structures \cite{Grabert}.  Recently, a great deal of
attention has been focused on a new class of nanometer-scale systems:
carbon nanotubes \cite{Ijima}.  The conducting properties of these
nanotubes, which are essentially tubular sheets of graphite, are predicted
to depend upon the diameter and helicity of the tube, parameterized by a
rollup vector (n,m).  One type of tube, the so-called (n,n), or armchair,
tube is predicted to be a novel 1D conductor with current carried by a
pair of one dimensional subbands \cite{Saito}, whose dispersion relations
near $E_f$ are indicated in the right inset to Figure 1.  A recent
breakthrough has made it possible to get large quantities of one
particular armchair nanotube, the (10,10) tube, which is approximately 1.4
nm in diameter \cite{Thess}.  This, coupled with recent successes in
performing electrical measurements on individual multiwalled nanotubes
\cite{Langer,Dai,Ebbeson,Kasumov} and nanotube bundles \cite{Fischer},
makes possible the study of the electrical properties of this novel 1D
system.

In this Report, we describe our measurements of transport through bundles,
or ropes, of nanotubes bridging contacts separated by 200-500 nm. We find
that a gap (suppressed conductance at low bias) is observed in the $I-V$
curves at low temperatures.  Further, dramatic peaks are observed in the
conductance as a function of a gate voltage $V_g$ that modulates the
charge per unit length of the tubes.  These observations are consistent
with single electron transport through a segment of a single tube with a
typical addition energy of $\sim$10 meV and an average level spacing of
$\sim$3 meV.

The device geometry is shown in the inset to Figure 1.  It consists of a
single nanotube rope to which lithographically defined leads have been
attached.  The tubes are fabricated as described in Ref. \cite{Thess} and
consist of ropes made up of ~1.4 nm diameter single-walled tubes. 
Nanodiffraction studies \cite{Bernaerts} indicate that approximately 30 -
40 $\%$ of these are (10,10) tubes.  Contacts are made to individual ropes
as follows.  First, the nanotube material is ultrasonically dispersed in
acetone and then dried onto an oxidized Si wafer upon which alignment
marks have been previously defined.  An atomic force microscope (AFM)
operating in the tapping mode is used to image the nanotubes.  Once a
suitable rope is found, its position is noted relative to the alignment
marks.  Resist is then spun over the sample, and electron beam lithography
is used to define the lead geometry.  A metal evaporation of 3 nm Cr/500
nm Au followed by liftoff forms the leads.  An AFM image of a completed
device is shown in the left inset to Figure 1. This device has four
contacts, allowing different segments of the rope to be measured and
four-terminal measurements to be performed.  The rope is clearly seen
underneath the metal layer, although it is not visible in between the
contacts due to the contrast of the image.  The device is mounted on a
standard chip carrier, contacts are wire bonded, and the device is loaded
into a He$^4$ cryostat. A dc bias can be applied to the chip carrier base
to which the sample is attached. This gate voltage $V_g$ modifies the
charge density along the length of the rope.  A number of samples have
been studied. All of the data presented here, however, were measured on a
single 12 nm diameter rope which should contain roughly 60 single-walled
nanotubes.

Figure 1 shows the current-voltage characteristics of the nanotube rope
section between contacts 2 and 3 as a function of temperature.  The most
notable feature is the appearance of a strong suppression of conductance
near $V = 0$ for temperatures below 10 K.  Gaps of a similar magnitude
were obtained for a number of nanotube ropes with diameters varying from 7
nm to 12 nm, and lengths from 200 to 500 nm.  There was no clear trend in
the size of the gap or the high-bias conductance with the rope length or
diameter.  We note that measurements of multi-walled nanotubes by us and
others \cite{Langer,Ebbeson,Kasumov} displayed no such gap in their $I-V$
curves.  These results are in rough agreement, however, with those
reported previously by Fischer et al. \cite{Fischer} and on similar, but
longer, ropes of single-walled nanotubes.  In their experiments, the
linear-response conductance also decreased at low temperatures.

Figure 2 shows the linear response conductance $G$ of the rope segment as
a function of $V_g$, measured at a temperature $T = 1.3$ K.  Remarkably,
the conductance consists of a series of sharp peaks separated by regions
of very low conductance.  The separation between peaks varies
significantly, but has a typical period of $\sim$ 1.5 V.  The peak
amplitudes vary widely, with the maximum amplitude of isolated peaks
approaching $e^2/h$.  The peaks are reproducible, although sudden changes
("switching") in the positions sometimes occur, particularly at larger
voltages.  The left inset to Figure 2 shows the temperature dependence of
a selected peak.  The peak width increases linearly with increasing $T$
(Figure 2: right inset) and the peak amplitude decreases.  The most
isolated peaks remain discernible even up to 50 K.

Figure 3 shows the differential conductance $dI/dV$ as a function of both
$V$ and $V_g$ for the rope segment between contacts 2 and 3.  The data are
plotted as an inverted grayscale, with dark corresponding to large
$dI/dV$.  The linear response peaks in $G$ (e.g. point A in the figure)
correspond to the centers of the crosses along the horizontal line at $V =
0$.  The gap in $dI/dV$ corresponds to the white diamond shaped regions
between the crosses (such as the region containing point B). These crosses
delineate the point of the onset of conduction at finite V (point C). 
Since the application of large biases led to significant switching of the
device, our sweeps were limited to ± 8 mV, and only the centers of the
diamond regions are visible.  Additional features (point D) are also
observed above the gap.

We now discuss our interpretation of these results.  We note that they are
very reminiscent of previous measurements of Coulomb blockade transport in
metal and semiconductor wires and dots \cite{Grabert}.  In these systems,
transport occurs by tunneling through an isolated segment of the conductor
or dot that is delineated either by lithographic patterning or disorder. 
Tunneling on or off this dot is governed by the single-electron addition
and excitation energies for this small system.  The period of the peaks in
gate voltage, $\Delta V_g$, is determined by the energy to add an
additional electron to the dot.  In the simplest model that takes into
account both Coulomb interactions and energy-level quantization, which we
refer to as the Coulomb blockade (CB) model, the peak spacing is given by: 
\begin{equation}
\Delta V_g= (U+\Delta E)/e \alpha                    
\end{equation}	
where $U=e^2/C$ is the Coulomb charging energy for adding an electron to
the dot, $\Delta E$ is the single-particle level spacing, and $\alpha=
C_g/C$ is the rate at which the voltage applied to the substrate changes
the electrostatic potential of the dot.  In the above, $C$ is the total
capacitance of the dot and $C_g$ is the capacitance between the back gate
and the dot.

To understand the dependence on $V$ and $V_g$ in more detail, consider the
energy level diagrams for the dot shown in the lower part of Figure 3. 
They show a dot filled with $N$ electrons, followed by a gap $U + \Delta
E$ for adding the $(N+1)th$ electron.  Above this, additional levels,
separated by the $\Delta E$, are shown, which correspond to adding the
$(N+1)th$ electron to one of the excited single-particle states of the
dot.  At a gate voltage corresponding to a Coulomb peak, the
electrochemical potential of the lowest empty state aligns with the Fermi
energy of the leads and single electrons can tunnel on and off the dot at
$V = 0$ (Figure 3A).  At gate voltages in between peaks (Figure 3B),
tunneling is suppressed due to the single electron charging energy $U$. 
As $V$ is increased so that the electrochemical potential of the right
lead is pulled below the energy of the highest filled state, an electron
can tunnel off the dot, resulting in a peak in $dI/dV$ (Figure 3C). 
Further increasing $V$ allows tunneling out of additional states, giving
additional peaks in $dI/dV$ (Figure 3D). Similar processes occur for
negative bias, corresponding to tunneling through unoccupied states above
the Coulomb gap.  At its largest, the required threshold voltage for the
onset of conduction of either type is: 
\begin{equation}
V_{max}=U+\Delta E.
\end{equation}

To apply this model to our system, we must postulate that transport along
the rope is dominated by single electron charging of a small region of the
rope or perhaps a single tube within the rope.  We will return to this
issue later.  For now, we will use the Coulomb blockade model to infer the
properties of this isolated region.  Further, we initially restrict
ourselves to the data of Figures 2 and 3, which corresponds to the central
rope segment.

The temperature dependence of Coulomb blockade oscillations can be used to
infer the parameters in equation (1).  In the CB model, the width of the
Coulomb oscillation is given by:  $d(\Delta V_g)/dT = 3.5 k_B/\alpha e$.
Comparison with the data in the right inset to Figure 2 gives $\alpha =
0.01$.  From this, and the measured spacing between peaks of 1-2 V, we
infer a typical addition energy: $U+ \Delta E = 10-20$ meV.  Also note
that the disappearance of the oscillations above $\sim$ 50 K yields a
similar estimate for the addition energy.

The amplitude of the conductance peak increases with decreasing
temperature at low temperatures.  Within the extended CB model, this
indicates that $\Delta E >> k_B T$ and that transport through the dot by
resonant tunneling though a single quantum level.  The peak height
decreases as $T$ is increased for temperatures up to roughly 10 K.  This
sets a lower bound on the energy level splitting of $\Delta E \sim 1$ meV. 
In addition for some peaks, such as those in the center of Figure 2, the
intrinsic linewidths of the peaks are clearly observable.  Fitting the
peak shapes (not shown) reveals that they are approximately Lorentzian, as
expected for resonant tunneling through a single quantum level
\cite{Grabert}.

The nonlinear $I-V$ measurements confirm the addition and excitation
energies given above.  The maximum size of the Coulomb gap $V_{max}$ in
Figure 3 is a direct measure of the addition energy - for the two peaks in
the figure, it is $\sim$ 14 meV. Tunneling through excited states is also
visible above the Coulomb gap for some peaks, and the level spacing to the
first excited state is found to range between 1-5 meV \cite{spacing}.  For
example, in Figure 3, the level spacing between states labeled by C and D
is $\Delta E = 1$ meV.

We now discuss how these parameters compare with expectations.  Consider a
single (n,n) nanotube.  The tube is predicted to be metallic \cite{Saito},
with two 1D subbands occupied at $E_f$ The order of magnitude of the
average level spacing should be related to the dispersion $dE/dk$ at the
Fermi level \cite{Saito,Zhu}: 
\begin{equation}
\Delta E \sim \frac{dE}{dk} \frac{\Delta k}{2} \sim \frac{dE}{dk} \frac{\pi}{L} \sim 
\frac{0.5 eV}{L [nm]} ,
\end{equation}
where the 2 arises from non-degeneracy of the two 1D subbands (see Fig. 1,
right inset.) The charging energy is more difficult to estimate
accurately.  The actual capacitance of the dot depends on the presence of
the leads, the dielectric constant of the substrate, and the detailed
dielectric response of the rope \cite{Benedict}. For an order of magnitude
estimate, however, we take the capacitance to be given by the size of the
object, $C = L$.  We then have: 
\begin{equation}
U=\frac{e^2}{C}=\frac{e^2}{L}=\frac{1.4 eV}{L[nm]} .
\end{equation}

Note the remarkable result that in 1D, both (3) and (4) scale like $1/L$,
and hence the ratio of the charging energy to the level spacing is roughly
independent of length.  This means that the level spacing will be
important even in fairly large dots, unlike in 3D systems. Using a length
of tube $L \sim$ 200 nm (the spacing between the leads) gives $U = 7$ meV
and $\Delta E = 2.5$ meV.  These are consistent with the observed values.
To relate these theoretical results for a single tube to the measurements
of rope samples, we first note that current in the rope likely flows along
a filamentary pathway \cite{nonlocal} consisting of a limited number of
single tubes or few-tube segments.  This is because, first, 60-70$\%$ of
the tubes are not (10,10), and hence the majority of the tubes in the rope
will be insulating at low $T$ \cite{intertube}. Second, the inter-tube
conductance is small compared to the conductance along the tube,
inhibiting inter-tube transport.  Finally, the metal likely only makes
good contact to the metallic tubes which are on the surface of the rope,
further limiting the number of tubes involved in transport.

Disorder along a filamentary pathway will effectively break it into weakly
coupled localized regions.  This disorder may result from defects
\cite{Chico}, twists \cite{Kane}, or places where inter-tube hopping is
necessary along the pathway.  Generally, the conductance would be expected
to be determined by single electron charging and tunneling between a few
such localized regions.  For other rope segments that we have measured,
the characteristics are consistent with transport through a few segments
in series/parallel, each with different charging energies.  For the
particular rope segment we have focused on here, however, a single
well-defined set of Coulomb peaks is observed, indicating that transport
is dominated by a single localized region.  Further, since the amplitude
of isolated peaks approaches the theoretical maximum for single-electron
transport of $e^2/h$, there is no other significant resistance along the
pathway.  We believe that this single region dominating transport is a
section of a single tube or at most a few-tube bundle.

While the above interpretation explains the major features in the data,
many interesting aspects of this system remain to be explored.  First one
would like to establish absolutely that transport is indeed occurring
predominantly along a single tube.  Second, it should be determined
whether all details of the data can be explained within the simple Coulomb
blockade model discussed above, since Coulomb interactions may
significantly modify the low-energy states from simple 1D non-interacting
levels \cite{Krotov}.  Of great interest would be measurements of
disorder-free tubes, where the intrinsic conducting properties of the tube
can be measured without the complications of single-electron charging. To
address these issues, experiments on individual single-walled tubes are
highly desirable, and both our group and others \cite{Tans} are making
progress in this direction.  Yet another important experiment would be to
measure directly the inter-tube coupling by making separate electrical
contact to two adjacent tubes.

In conclusion, we have measured transport in single nanotube ropes at low
temperatures.  We find that a Coulomb gap appears in the $I-V$, Coulomb
oscillations are observed as a function of an external gate voltage, and
structure is observed in the $I-V$ curve above the Coulomb gap.  We
interpret these results as evidence for single-electron transport through
the quantized energy levels of the nanotubes comprising the rope. 

We wish to acknowledge useful discussions with Vincent Crespi, Marvin
Cohen, Dung-Hai Lee, and Steven Louie.  We are also indebted to Sander
Tans for communicating his unpublished results on measurements on
nanotubes and emphasizing the importance of a gate voltage in modulating
the properties. This work was supported by the U.S. Department of Energy
under Contract No. DE-AC03-76SF00098 and by the Office of Naval Research,
Order No. N00014-95-F-0099. 


\begin{figure}
\caption[]{Left inset: Atomic Force Microscope (AFM) micrograph of a 
completed device. The bright regions are the lithographically defined 
metallic contacts, labeled 1-4. The rope is clearly visible as a 
brighter stripe underneath the metallic contacts.  In between the 
contacts (dark region), it is difficult to see the rope because of the 
image contrast.  Note that the width of the rope in the AFM image 
reflects the convolution of its actual width with the AFM tip radius 
of curvature.  The actual thickness of the rope is experimentally 
determined by the measuring its height with the AFM and assuming that 
the rope is cylindrical.  Main:  The $I-V$ characteristics at different 
temperatures for the rope segment between contacts 2 and 3.   Right 
inset:  Schematic energy-level diagram of the two one-dimensional 
subbands near one of the two Dirac points [3], with the quantized energy 
levels indicated. The k-vector here points along the tube axis.}
\label{fig1}
\end{figure}

\begin{figure}
\caption[]{Conductance versus gate voltage at $T =1.3$ K for the segment 
between contacts 2 and 3.   Left inset:  Temperature dependence of  a 
peak. Note that this peak was measured on a different run from the data 
in the main panel and does not directly correspond to any of those peaks.  
Right inset:  Width of the peak in the left panel as a function of 
temperature.}
\label{fig2}
\end{figure}

\begin{figure}
\caption[]{Top: gray scale plot of the differential conductance  $dI/dV$ 
of the rope segment between contacts 2 and 3, plotted as a function of 
$V$ and $V_g$.  To enhance the contrast of the image, a smoothed version 
of the data was subtracted from the differential conductance. Bottom:  
Schematic energy-level diagrams of the device within the Coulomb blockade 
model at the points indicated in the upper panel. Panel A shows the 
linear-response ($V = 0$) transport possible at a Coulomb peak. Panel B 
shows the blockaded regime between peaks; indicated are the addition 
energy $U+\Delta E$ and the level spacing $\Delta E$.  Panels C and D 
show the diagram at two different applied voltages corresponding to 
transport through the first and second occupied states, respectively.}
\label{fig3}
\end{figure}

\widetext

\begin{thebibliography}{10}
\bibitem{spacing} The level spacing can be determined from the data in
Fig. 3 by two means.  The first is to measure the separation between
features in $V_g$, and employ the conversion factor $\alpha =
V_{max}/\Delta V_g$.  The second is to measure the separation between the
features in $V$, and use the slopes of the lines defining the Coulomb gap
to infer $\alpha_L = C_L /C$ and $\alpha_R = C_R /C$.  For more
information, see Ref \cite{Grabert}.
\bibitem{nonlocal} Other measurements on these samples also support the
notion that transport is through individual decoupled tubes.  In four
terminal measurements of the ropes at temperatures $> 30$ K (where the
rope resistance becomes much less than the voltage probe impedance and
reliable four-terminal measurements are possible), non-local voltages have
been observed.  For example, referring to the inset to Figure 1, a voltage
can appear between the terminals 1 and 2 in response to a current flowing
between contacts 3 and 4.  Such non-local effects are possible if
different contacts make contact to different tubes and the inter-tube
coupling is small.  Current injected into 4 in a particular tube that is
not well connected to 3 can arrive at contact 1 or 2, producing a
non-local response.
\bibitem{intertube} Individual tubes within a rope have chiral angles
within 10 degrees of the achiral (10,10) tube with roughly 30-40$\% $ of
the sample being (10,10) tubes \cite{Bernaerts}. The indices of tubes
consistent with the experimental constraints on chirality and radius
\cite{Thess} are (10,10), (9,11), (8,12), (7,13) and the opposite-handed
twins. The gap of the (7,13) tube is small and likely does not survive
intertube interactions, but the gaps of the (9,11) and (8,12) tubes are
about 0.5 eV, which is probably large enough to maintain semiconducting
behavior within the rope.
\end{thebibliography}
\end{document}